\documentclass[12pt,a4paper,]{article}
\usepackage[latin1]{inputenc}
\usepackage{amsmath}
\usepackage{amsfonts}
\usepackage{indentfirst}
\usepackage{amssymb}
\usepackage{graphicx}
\usepackage{mathptmx}
\usepackage{color}
\usepackage[super,comma,sort&compress]{natbib}
\usepackage[left=2.00cm, right=2.00cm, top=2.00cm, bottom=2.00cm]{geometry}
\begin{document}
	\linespread{2.0}
	\selectfont
	
\begin{center}
	{\LARGE \textbf{Role of Charge Traps in the Performance of Atomically-Thin Transistors}}\\ \ \\
	Iddo Amit\textsuperscript{1}, Tobias J. Octon\textsuperscript{2}, Nicola J. Townsend\textsuperscript{1}, 
	Francesco Reale\textsuperscript{3}
	, C. David Wright\textsuperscript{2}, 
	Cecilia Mattevi\textsuperscript{3}
	, Monica F. Craciun\textsuperscript{2} and Saverio Russo\textsuperscript{1}$^{\dagger}$\\ \ \\
	\textsuperscript{1}Centre for Graphene Science, Department of Physics, University of Exeter, Stocker Road, Exeter, United Kingdom, EX4 4QL\\
	\textsuperscript{2}Centre for Graphene Science, Department of Engineering, University of Exeter, North Park Road, Exeter, United Kingdom, EX4 4QF\\
	\textsuperscript{3}Department of Materials, Imperial College London, United Kingdom, SW7 2AZ\\
	$^{\dagger}$\texttt{S.Russo@exeter.ac.uk}
	
\end{center}
\section*{Abstract}
Transient currents in atomically thin MoTe$_2$ field-effect transistor are measured during cycles of pulses through the gate electrode. The transients are analyzed in light of a newly proposed model for charge trapping dynamics that renders a time-dependent change in threshold voltage the dominant effect on the channel hysteretic behavior over emission currents from the charge traps. The proposed model is expected to be instrumental in understanding the fundamental physics that governs the performance of atomically thin FETs and is applicable to the entire class of atomically thin-based devices. Hence, the model is vital to the intelligent design of fast and highly efficient opto-electronic devices.

\newpage
\setlength{\parskip}{2ex}
\linespread{2.0}
\selectfont

The emerging family of atomically thin materials is fueling the development of conceptually new technologies\cite{Splendiani2010Emerging} in highly-efficient optoelectronics\cite{Wang2012Electronics,Chen2016Nanostructured} and photonic applications,\cite{Xia2014Two} to name a few. The large variety of band gap values found in layered transition metal dichalcogenides (TMDC)\cite{Tang2013Graphene,Fiori2014Electronics} make these materials especially suited for transistor applications. TMDCs are compounds with the general formula MX$_2$, where M is a transition metal, \textit{e.g.} Mo and W, and X is an element of the chalcogen group, S, Se and Te. They appear in a layered structure where the metal forms a hexagonal plane and the chalcogenides are positioned over and under this plane in either a trigonal prismatic (2H), as shown in Fig. 1(a), or octahedral (1T) stacking configuration.\cite{Lv2014Transition} In the semiconducting 2H systems, the compounds show a transition from indirect band gap in bulk materials to direct band gap in single layers.\cite{Mak2010Atomically} 

Single and few-layers TMDCs have been implemented in a wide range of applications, ranging from thin film transistors,\cite{Das2014All} digital electronics and opto-electronics,\cite{Jariwala2014Emerging,Wang2012Electronics,Das2014High} flexible electronics,\cite{Akinwande2014Two} and up to energy conversion and storage devices.\cite{Wang2014Graphene} However, the defect states in TMDCs have an ambivalent nature and can have a major positive or negative impact on the performance of atomically-thin devices. The presence of defects in photodetectors can be beneficial since it has been shown to immobilize charges at the channel which improves the gain in photodetectors\cite{Furchi2014Mechanisms} and produces non-volatile memory mechanisms.\cite{Lee2012Mos2} On the other hand, large hysteresis caused, for example, by charge traps\cite{Wang2012Electronics} and significant Schottky barriers\cite{Allain2015Electrical} at the metal-semiconductor interface are still a major design challenge for the realisation of novel device architectures. They have been shown to cause degradation in the performance of transistors\cite{Pu2016Highly} and generate high levels of flicker noise.\cite{Cho2015Low,Cho2015Temperature} To overcome these challenges, hysteresis is usually avoided by encapsulation\cite{Lee2013Flexible,Kufer2015Highly} or operation under high-vacuum.\cite{Jariwala2013Band,Sangwan2013Low}

Most of the current research into surface states of TMDCs has focused on the chemical origins of charge trapping. A full understanding of their effect on the electrical properties is still lacking, hindering the optimization of functional components. While hysteresis has been shown to correlate with traps generated at the channel-dielectric interface and the channel-ambient interface,\cite{Park2016Thermally,Late2012Hysteresis} little attention has been given to the mechanisms by which immobile charges affect the conduction characteristics of the devices, which is fundamentally different from those experienced in bulk devices. 

In this communication we present the first study of the role of immobile charges on the electrical transport properties of atomically thin MoTe$_2$. This TMDC is of particular interest since its direct band gap of 1eV\cite{Lezama2014Surface,lezama2015Indirect} matches the wavelength of maximal solar emission intensity, thus making it a prime candidate for solar energy converters. MoTe$_2$ is intrinsically \textit{p}-doped, but can exhibit ambipolar behavior,\cite{Xu2015Reconfigurable,Lezama2014Surface} mobility in the range of 10--30 cm$^2$ V$^{-1}$s$^{-1}$,\cite{Lezama2014Surface,Pradhan2014Field} and on-off ratios of up to $10^6$.\cite{Pradhan2014Field} A stringent quantitative analysis demonstrates that the role of trapped charges in the operation of MoTe$_2$-based electronic components is a change in the threshold voltage of the field-effect transistor (FET), effectively modulating the resistivity of the entire channel. By repeating the charge capture and emission cycles in different drain biases we are able to distinguish between two sources of transient behavior in MoTe$_2$-FETs. One transient is due to emission of charges from traps to the channel, and the other is due to time-dependent capacitive gating of the channel that produces a transient in the effective threshold voltage. Finally, we present a complete analytical model to support our observations. Our findings are applicable to the entire class of atomically thin-based devices and provide a thorough understanding of charge traps and carrier dynamics which is needed to facilitate the intelligent design of fast and highly efficient opto-electronic devices.


Few-layers MoTe$_2$ flakes were obtained by mechanical exfoliation of 2H-MoTe$_2$ bulk crystal (\textit{HQ graphene}) onto highly doped silicon substrates, covered with 290 nm of high-quality thermally grown SiO$_2$. The silicon substrate was used as a global back gate electrode, with the oxide layer acting as a gate dielectric. Standard electron beam lithography procedure was used to pattern electrodes and electrical leads. The contacts were then immediately metalized with 5 nm of Ti adhesion layer, and 50 nm of Au, using an electron beam evaporation system, working at very low pressure ($\sim 10^{-8}$ mbar) and at long working distance, to achieve high uniformity in the deposition. The devices were then annealed in dry Ar/H$_2$ environment at ambient pressure for 2 hours at 200\textdegree C. Fig. 1(b) shows a schematic representation, not to scale, of the device and the circuit details. Atomic force microscopy measurements (Fig. 1(c)), and optical contrast (not shown here) of the flakes confirm that the surface of MoTe$_2$ is not visibly contaminated and that the studied flakes consists of 4 number of layers.

\begin{figure}
	\centering
	\includegraphics[scale=0.25]{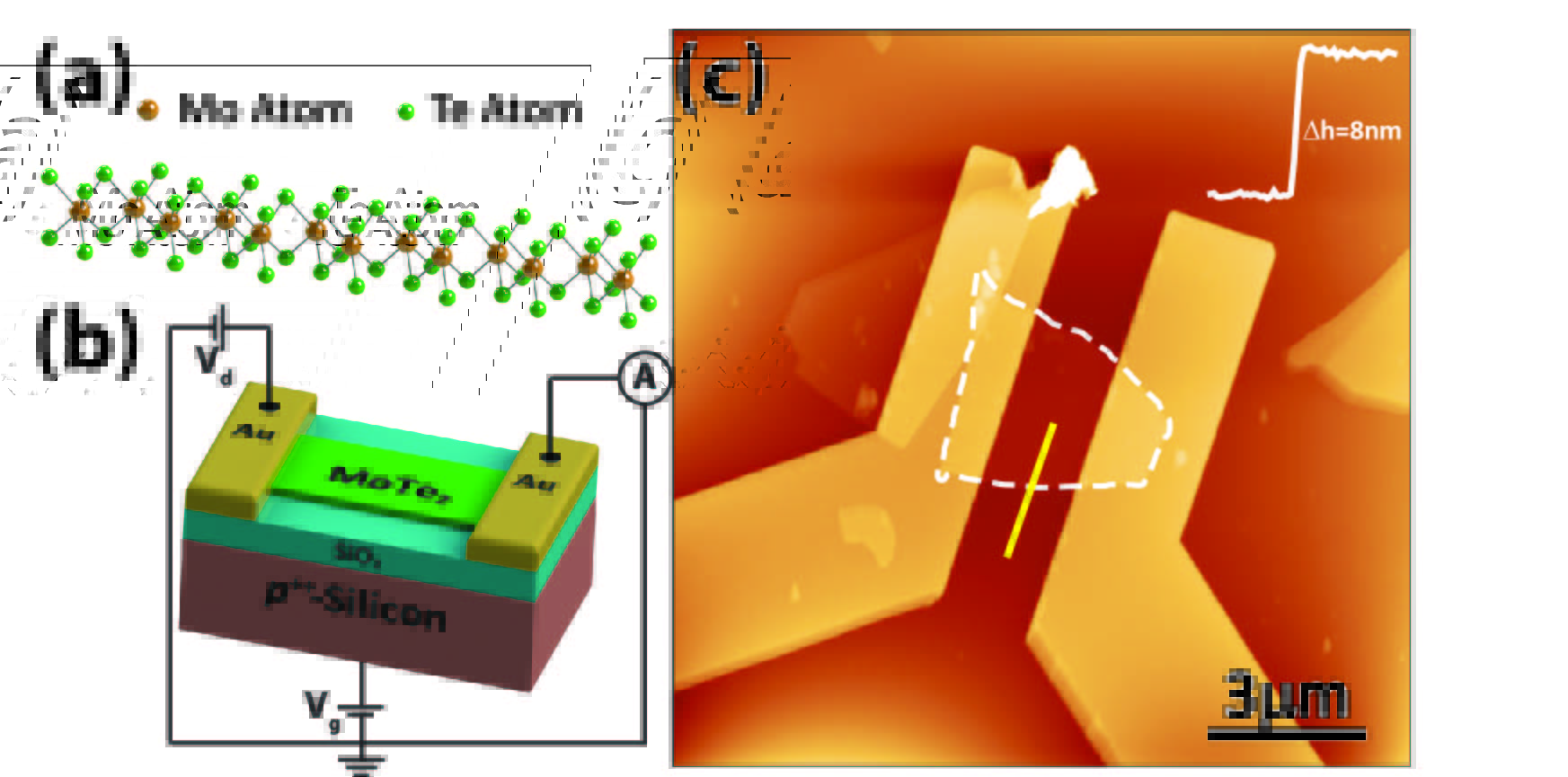}
	\caption{Panel (a) shows a three dimensional model of the 2H-MoTe$_2$ crystal structure, with a single layer of the trigonal prismatic stack. Panel (b) shows the schematics of the device architecture and measurement setup. Panel (c) is atomic force microscope image of a typical device, showing the source and drain symmetric electrodes and the MoTe$_2$ flake (outlined in dashed white line). The inset shows a scan profile (taken along the yellow line) from the substrate to the flake.}
\end{figure}

Low noise electrical measurements were performed in a home-built Farady cage in the dark and in ambient conditions on more than five different devices, all showing a similar behavior. The drain electrode was biased using a low noise voltage source and the source electrode was kept grounded throughout the experiment. The current flowing through the source electrode was measured using a current preamplifier. An independent voltage source-meter was used to apply a bias to the gate electrode while measuring the leakage current. The response time of the system was found to be limited only by the minimal rise time of the preamplifier, which is $<5\mu s$. (See Supporting Information) 


The electronic behavior of multiple devices was characterized by measuring their drain current \textit{vs.} voltage response (I$_{\text{ds}}$-V$_{\text{ds}}$) and drain current \textit{vs.} gate voltage transfer (I$_{\text{ds}}$-V$_{\text{gs}}$) characteristics. Figure 2(a) shows the response curve of a typical MoTe$_2$ transistor. The curve exhibits a slight asymmetry with higher resistivity for negative applied drain bias, indicating that the metal-semiconductor contacts form a small Schottky barrier for holes. The origin of this asymmetry about V$_{\text{ds}} = 0$V is in the different electrostatic potential seen by the source and drain electrodes. In the experiment the potential barrier at the MoTe$_2$/source electrode interface is kept constant, as it is pinned by the gate. On the other hand, the biased drain barrier decreases (increases) in height with positive (negative) drain bias.\cite{Tian2014Novel} Despite the low Schottky barrier, both the linear and the log-scale of the response curve (inset in Fig 2(a)) show that the device is not rectifying and is, in fact, largely Ohmic in higher V$_{\text{ds}}$ values (see Supporting Information).

\begin{figure}
	\centering
	\includegraphics[scale=1]{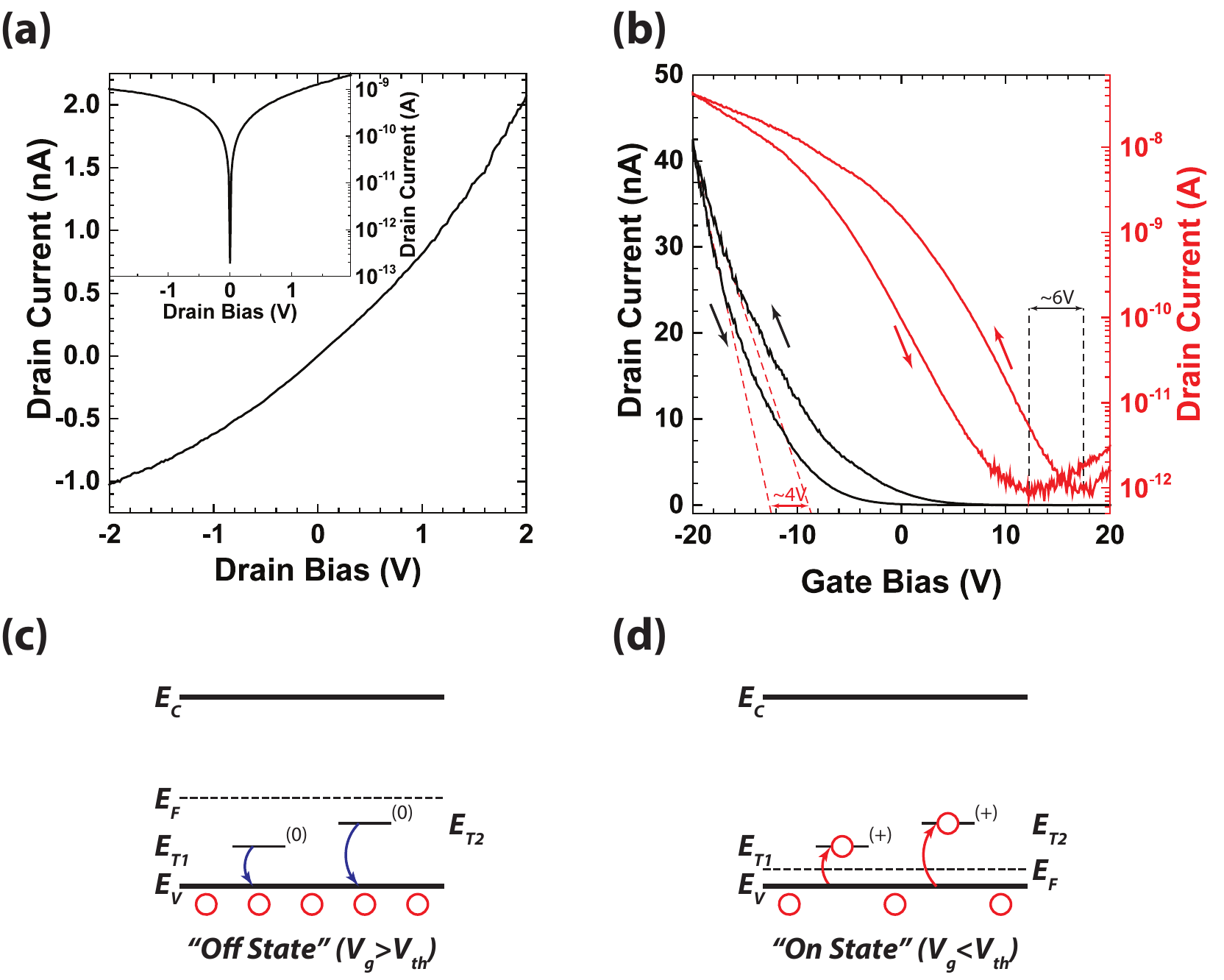}
	\caption{Panel (a) shows the response (I$_{\text{ds}}$-V$_{\text{ds}}$) curve of a typical field effect transistor, taken with zero gate bias (V$_{\text{gs}} = 0$). The inset shows the same curve in a semi-logarithmic scale. Panel (b) shows the transfer (I$_{\text{ds}}$-V$_{\text{gs}}$) curve of the same device taken with 1 V source drain bias (V$_{\text{ds}}$) shown in a linear (solid black) and semi-logarithmic (solid red) scale. The dashed red lines are a linear extrapolation of the linear part of the curve, showing a change of 4 V in threshold voltage. The dashed black lines indicate the change in the position of the charge neutrality point. The arrows indicate the back gate sweep direction. Panels (c) and (d) show schematic energy band diagrams for the emission (c) and capture process (d) when the channel is in the ``off state'' and ``on state'', respectively. $E_C$, $E_V$, $E_F$, $E_{T1}$ and $E_{T2}$ are the conduction band minimum, the valance band maximum, the Fermi energy, the shallow midgap state and deep midgap state energy, respectively.}
\end{figure}

The device transfer characteristics are shown in Fig. 2(b), taken at V$_{\text{ds}} = 1$V. The curve matches the expected behavior of an enhancement-mode \textit{p}-channel transistor, showing an increase in drain current as the gate bias grows more negative beyond the threshold voltage (V$_{\text{th}}$). From the transfer curve, we can estimate the device mobility, $\mu _p$, and sub-threshold swing, $SS$. Using $\mu _p = L (d\text{I}_{\text{ds}}/d\text{V}_{\text{gs}})/(WC_{ox} \text{V}_{\text{ds}})$ in the linear regime of the curve, where $L = 1\mu$m and $W = 3\mu$m are the device length and width, respectively, and $C_{ox} = \varepsilon _0 \varepsilon _r / d = 115 \mu$F m$^{-2}$ is the gate dielectric capacitance, with $\varepsilon _0$ the vacuum permittivity and $\varepsilon _r$ the oxide relative permittivity, we find that the mobility is between 0.12 on the forward sweep and 0.14 cm$^2$V$^{-1}$s$^{-1}$ on the back sweep. From the sub-threshold part of the curve, we estimate a sub-threshold swing value of 4 V dec$^{-1}$ using $SS = \left(d\log _{10} \text{I}_{\text{d}}/d\text{V}_{\text{g}}\right)^{-1}$. The low value of the mobility and the high value of the swing are indicative of the presence of mid-gap trap states.\cite{Furchi2014Mechanisms}. In line with these findings, the gate sweep measurements also show a hysteretic behavior resulting in a shift in V$_{\text{th}}$ between the forward and backward sweeps, which changes the threshold voltage by about $\Delta$V${\text{th}} = -4$V and the charge neutrality point by about $-6$V, see Fig. 2(b). 

To understand the physical origin of the observed changes in threshold voltage we use the well-known equation that describes V$_{\text{th}}$ in field-effect transistors:
\begin{equation}\label{VTH}
\text{V}_{\text{th}} = \Phi _{MS} - \frac{Q_i}{C_{ox}} - \frac{Q_T}{C_{ox}} -\Delta E_F
\end{equation}  
where $\Phi _{MS}$ is the difference between the metal and semiconductor workfunctions when all the terminals are grounded, $C_{ox}$ is the gate dielectric capacitance, $Q_i$ is the static charge density within the dielectric, $Q_T$ is the trapped charge density at the interface between the dielectric and the conductive channel and $\Delta E_F$ is the shift in the Fermi Energy, required to turn the transistor on. From Eq. \ref{VTH} it is clear that the only parameter that can change during the back gate sweep is the population of midgap traps, $Q_T$, indicating that positive charges (holes) are immobilized during the sweep. The process of charge trapping is illustrated in the energy band diagrams of Fig. 2(c+d) using two "donor-type" mid-gap states. In the ``off state'', where the Fermi level is above the trap levels ($E_{T1}$ and $E_{T2}$) the traps are occupied by an electron and are neutral. In the ``on state'', the traps are void of electrons (occupied by a hole) and are positively charged. 

\textit{A priori}, the observed hysteresis can be due to charge trapping in the metal-semiconductor interface, \textit{i.e.} localized at the contacts region, or at the entire surface area of the channel, \textit{i.e.} at the semiconductor-dielectric and -ambient interface. However, the changes in the transfer curve strongly suggest that most of the charge trapping occurs throughout the entire area of the conductive channel, rather than at the metal-semiconductor interface. The noticeable shift in the charge neutrality point with respect to the gate bias (minimal conductivity in the log-scale, red curve) in Fig. 2(b), is indicative of a change in effective doping of the channel due to the space charge region generated by the immobilized charges. In contrast, a change in the degree of Fermi-level pinning at the contacts would have manifested primarily in changes in the linear slope of the logarithmic curve (the sub-threshold slope) and by changes in the width of the neutrality point. Assuming that the trapped charges are distributed in the channel, an assumption that is further validated by the analysis of the threshold transients, we can estimate that the difference in trapped charge density between the forward and back sweep is about $4.3\times 10^{11}$cm$^{-2}$, using $\Delta Q_T = \Delta \text{V} C_{ox}$.

To gain insight on the dynamics of the charge traps, their effect on the transfer currents and their role in producing hysteretic cycles, we have monitored the transport characteristics while pulsing the gate electrode from ``open'' (more negative) to a ``close'' (more positive) value. The drain current was recorded over long periods of time (60-90 minutes) while the gate was repeatedly pulsed between V$_{\text{gs}} = -10$V to open the channel and V$_{\text{gs}} = 0$V to close it (Top panel in Fig. 3(a)). As the pulse on gate drives the channel from a close to an open state, a sudden rise of the current in the channel is measured followed by a fast decay. When the gate is pulsed back to the closed state, the current drops down and then slowly begins to recover. The decay in current in the open state is due to the capturing of holes in mid-gap traps that shifts the threshold voltage to a more negative value (red arrows in Fig. 2(d)), effectively closing the channel. On the other hand, the recovery in the off state is due to the holes that are emitted from the traps (blue arrow in Fig. 2(c)) shifting V$_{\text{th}}$ to a less negative value. While the capture process is spontaneous and fast, the emission mechanism is thermally activated and, therefore, significantly slower then the capture rates. 

\begin{figure}
	\centering
	\includegraphics[scale=0.225]{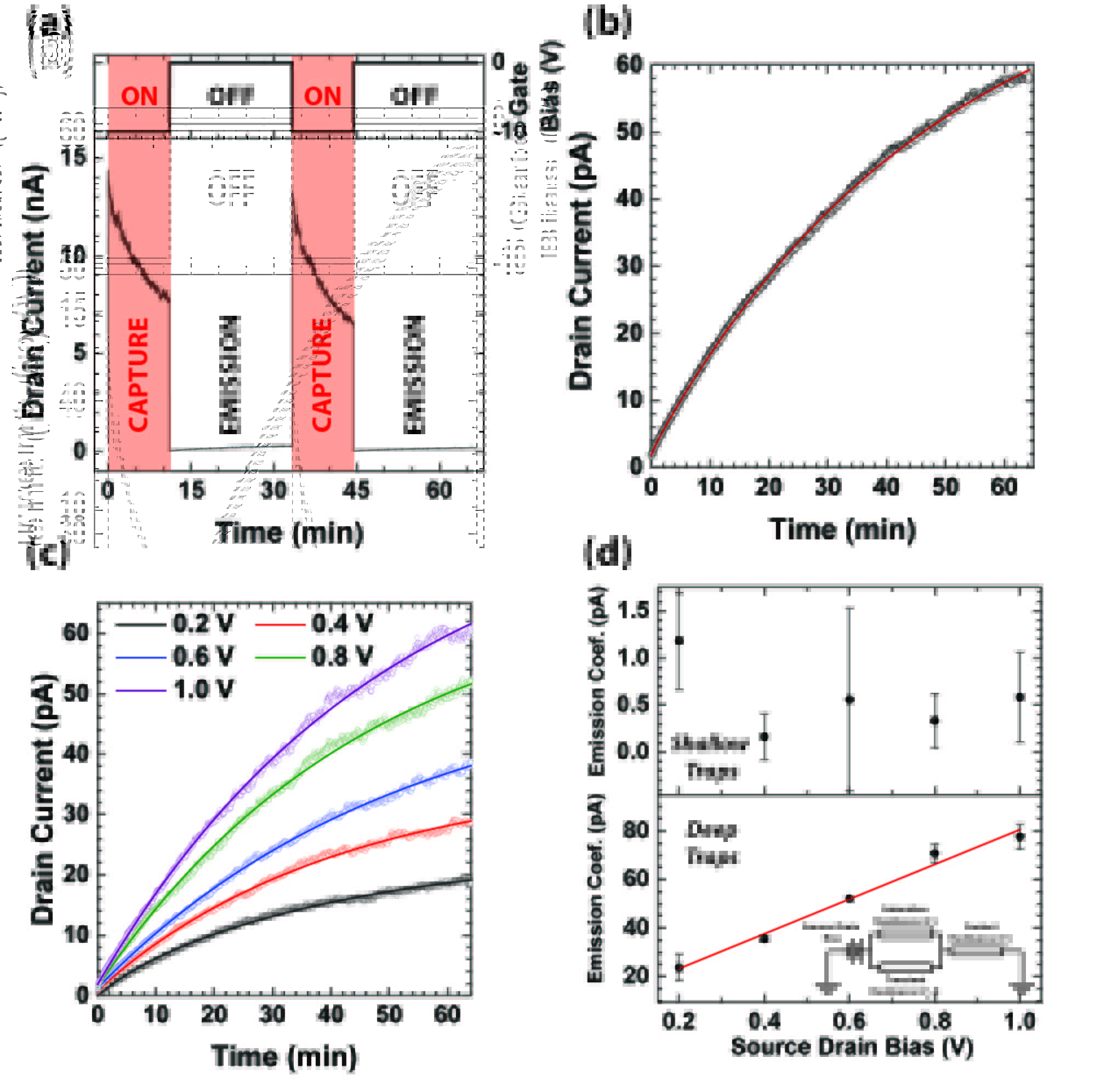}
	\caption{Panel (a) shows the gate pulse cycles. The ``On'' and ``Off'' segments are highlighted. On the top, the applied gate voltage during each segment. On the bottom Panel: The drain current during the capture (on red background) and emission (on white background) segments. (b) An emission segment, recorded at V$_{\text{gs}} = 0V$ and V$_{\text{ds}} = 1V$, averaged over four cycles. The black circles are the measured data and the red curve is the fit to a double-exponential rise equation. (c) Emission segments, recorded at V$_{\text{gs}} = 0V$ and varying V$_{\text{ds}}$ values, from 0.2V to 1.0V in 0.2V intervals. The circles are the measured data and the solid curves are the double-exponential fits. (d) The pre-exponential coefficients for the short emission coefficients, $A _1$ (Top Panel) and long emission coefficient $A _2$. The red line represents the best linear fit. Inset: An equivalent circuit diagram of the transient threshold model proposed here.}
\end{figure}

The vast majority of models used to quantify the time-dependent behavior of charge emission from mid-gap traps are based on Schottky or asymmetric diode structures.\cite{Lang1974Deep,Borsuk1980Current} These models accurately describe the currents, and the resulting transient changes in capacitance, that are associated solely with the emission of charges from traps back into the circuit. However, transient changes in threshold voltage should affect the measured current in a completely different way, which has not yet been studied though it plays a pivotal role for the development of fast opto-electronic applications. 

To elucidate the fundamental difference in transient behaviors, we must first describe the main aspects of the conventional semiconductor model for current transients. When the emission of charges from depletion regions takes place, the current has a constant (saturation) component, which is a function of the applied bias, and a transient component which is the emission current:
\begin{equation}\label{current_transient}
I(t) = I_0 + \frac{qN_TA}{\tau} e^{-t/\tau}
\end{equation}
$I_0$ is the saturation current, $q$ is the elementary charge and $A$ is the surface area of the device contact. Within this model, the time dependence of the transient current is a function of the density of trapped charges ($N_T$) and the decay coefficient $\tau$ which is a function of the energetic position of the trap with respect to the valance band.(see supporting information). However, in atomically thin MoTe$_2$ FET, the high sensitivity of the conducting channel to its surrounding media means that the charge carrier dynamics can lead to significant shifts in threshold voltage and charge neutrality point, effectively changing the resistance of the entire channel. Hence, an inclusive model in which the resistivity changes with time is needed. To this end, we use the well known expression that describes the linear regime of the transfer curve, where the current is determined by:\cite{Sze2006Physics}
\begin{equation}
\text{I}_{\text{d}}(t) = \frac{W\mu _p C_{ox}}{L}\left(\text{V}_{\text{th}} (t) - \text{V}_{\text{g}}\right) \text{V}_{\text{d}}
\end{equation}
Where $W$ and $L$ is the channel width and length, respectively, and $\mu _p$ is the hole mobility. Since in atomically-thin FETs, the only time-dependent component of the threshold voltage (Eq. \ref{VTH}) is the density of trapped charges we can write $d\text{V}_{\text{th}} (t)/dt = -(q/C_{ox}) \left(dp_T(t)/dt\right)$ where $p_T = Q_T/q$ is the density of occupied traps. To obtain a full description of the threshold voltage transient, V$_{\text{th}}(t)$, we assume that the density of free carriers, $p$, directly correlates to the equilibrium density $p_0$, by $p = p_0 - p_T$, \textit{i.e.} that there is no net injection of charges through the contacts. We further use the well-known result of the Shockley-Reed-Hall derivation to write the time-dependent density of occupied traps as $p_T (t) = N_T e^{-t/\tau}$. The transient of the threshold voltage then becomes:
\begin{equation}\label{VTHt}
\text{V}_{\text{th}} (t) = \text{V}_{{\text{th}},sat} - \frac{qN_T e^{-t/\tau}}{C_{ox}}
\end{equation}
where all the time-independent quantities have been grouped in V$_{\text{th},sat}$ for convenience. With the expression for V$_{\text{th}} (t)$ from Eq. \ref{VTHt}, the expression for the transient current is readily obtained:
\begin{equation}\label{ITH}
\text{I}_{\text{d}}(t) = \text{I}_{{\text{d}},sat} - \frac{q W\mu _p N_T \text{V}_{\text{d}}}{L} e^{-t/\tau} .
\end{equation}
The expression in Eq. \ref{ITH} has one striking difference from the conventional expression for current transient (Eq. \ref{current_transient}), it is linear with drain bias. Qualitatively, this is a simple manifestation of Ohm's law: as the resistance of the conductive channel changes with time, the current responds linearly, proportional to the applied bias.

In the emission segments of the gate-pulse experiment, 
we find that a significant increase in currents occurs on a very short time scales, while a further, slower increase is easily discernible in longer time scales. This behaviour cannot be satisfied by a single exponential fit but is in excellent agreement with a
double exponential rise equation in the form $I(t) = I_0 + A_1e^{-t/\tau _1} + A_2 e^{-t/\tau _2}$ (red line in Fig. 3(a)) suggesting that there are two types of traps\cite{Late2012Hysteresis}, a shallow trap and a deeper one, corresponding to emission coefficients $\tau _1 \approx 250s$ and $\tau _2 \approx 2,900s$. Fig. 3(b) shows the recovery currents, measured by pulsing the gate between -10 V and 0 V at drain bias values ranging from 0.2 V to 1 V. The curves are then fitted with a double exponential rise curve, without any assumption on the form of the pre-exponential factors, $A_1$ and $A_2$, while maintaining the emission constants within reasonable boundaries. 

To distinguish between the different contributions to the transient current, the pre-exponential factors of the shallow and deep traps are plotted in Fig. 3(d) on the top and bottom panel, respectively. Within the measurement error, it is clear that the pre-exponential factor of the transient current that is due to emission from the shallow traps is constant, and independent of the drain bias. This suggests that the measured signal is, indeed, the emission current from the traps. For the deep traps, the pre-exponential factors are found to have a linear dependence on V$_{\text{ds}}$. This is expected for deep traps that are uniformly distributed about the conductive channel and are not simply concentrated at the  metal-semiconductor interface, and is consistent with the analysis of the hysteresis of the gate bias measurements. Comparing the two panels in Fig. 3(d)  reveals two striking features in the transient mechanism. First, the two orders of magnitude difference in the pre-exponential coefficients show that the threshold transient is the significant factor, governing the transistor response over time. Second, the change in the trap population ($\Delta N_T = L(dA_2/d\text{V}_{\text{d}})/(qW\mu _p) \sim  10^9$ cm$^{-2}$) is a small fraction of the overall estimated density of $10^{12}$ states per cm$^{2}$,\cite{Lin2015Origin} corresponding to the small dynamic window of operation used here. This emphasizes the significant role that the threshold voltage transients play in the behavior of atomically thin MoTe$_2$ transistors.


The presented model of the threshold voltage transients is general, since it does not take into account features which are specific to MoTe$_2$. 
For example, similar studies conducted on WS$_2$ grown by chemical vapor deposition also show a bi-exponential decays of the transient current which is fully captured by our model (see Supporting Information).
Most importantly, this model is independent of the spatial location of trapped charge states (\textit{e.g.} semiconductor-substrate or -ambient interface) and it is universally valid for semiconductor channels thickness that are significantly smaller than the Debye screening length, a condition easily met in emerging atomically thin materials. Our proposed model of threshold voltage transients can be further expanded and included in well established methodology of charge trap spectroscopy, whether probed by temperature scans\cite{Chen2009Density} or by optical means.\cite{Lee2010Interfacial} However, the added simplicity of our methodology means that it can be applied to a variety of materials and substrates, including those that are photo-active, or temperature sensitive.

Finally, we calculate the overall resistance of the device and find that the transient resistance operates in parallel to the saturation resistance:
\begin{equation}
\left(\frac{d\text{I}_{\text{d}}(t)}{d\text{V}_{\text{d}}}\right) = \left(\frac{d\text{I}_{{\text{d}},sat}(t)}{d\text{V}_{\text{d}}}\right) - \frac{qW\mu _p N_T}{L}e^{-t/\tau}
\end{equation}
or $R^{-1} = R_{sat}^{-1} + R_{trans}^{-1}$ which is a strong indication to the fact that both factors indeed stem from the channel itself. We note that the addition of series resistance to the circuit, such as contact resistance, does not affect the time-dependent characteristics of the model, as is discussed in details in the Supporting Information.


In conclusion, we have demonstrated a new approach to the analysis of charge trapping and transient response of TMDC-based FETs, which paves the way to a better understanding of the role of mid-gap state in the operation novel devices. Using a simple two terminal model system, we were able to distinguish between currents associated with the emission of trapped charges into the circuit and currents that evolve in time due to the changes in effective threshold voltage across the channel. The mechanism of threshold voltage transients which we study and model is not limited to MoTe$_2$ but it is valid to any device based on atomically thin materials. Indeed, as long as the channel depth is much smaller than the Debye screening length, the threshold voltage will be strongly modulated by the formation of space charge regions at both the semiconductor-dielectric and -ambient interfaces. Our model, which describes the basic physics that govern the hysteretic characteristics of atomically thin FETs, is instrumental for the design of defect-based devices, such as photodetectors and memory devices, as well as provides a new methodology to study the nature of these defects.

\subsection*{Acknowledgments}
Iddo Amit acknowledges financial support from The European Commission Marie Curie Individual Fellowships (Grant number 701704). S. Russo and M.F. Craciun acknowledge financial support from EPSRC (Grant no. EP/J000396/1, EP/K017160/1, EP/K010050/1, EP/G036101/1, EP/M001024/1, EP/M002438/1), from Royal Society international Exchanges Scheme 2016/R1 and from The Leverhulme trust (grant title "Quantum Drums" and "Room temperature quantum electronics"). 
N.J. Townsend and S. Russo acknowledge DSTL grant scheme Sensing and Navigation using quantum 2.0 technologies. C.M. acknowledges the award of a Royal Society University Research Fellowship by the UK Royal Society, and the EPSRC-Royal Society Fellowship Engagement Grant EP/L003481/1.

\newpage
\bibliographystyle{AdvMatBst}
\bibliography{MoTe2bib}

\newpage
\linespread{2.0}
\selectfont

\begin{center}
	{\large \textit{Supporting Information}}:\\ \ \\
	{\LARGE \textbf{Transient Threshold Voltage in Atomically Thin MoTe$_2$ Transistors}}\\ \ \\
	Iddo Amit\textsuperscript{1}, Tobias J. Octon\textsuperscript{2}, Nicola J. Townsend\textsuperscript{1}, Francesco Reale\textsuperscript{3}, C. David Wright\textsuperscript{2}, Cecilia Mattevi\textsuperscript{3}, Monica F. Craciun\textsuperscript{2} and Saverio Russo\textsuperscript{1}$^{\dagger}$\\ \ \\
	\textsuperscript{1}Centre for Graphene Science, Department of Physics, University of Exeter, Stocker Road, Exeter, United Kingdom, EX4 4QL\\
	\textsuperscript{2}Centre for Graphene Science, Department of Engineering, University of Exeter, North Park Road, Exeter, United Kingdom, EX4 4QF\\
	\textsuperscript{3}Department of Materials, Imperial College London, United Kingdom, SW7 2AZ\\
	$^{\dagger}$\texttt{S.Russo@exeter.ac.uk}
\end{center}
\setlength{\parskip}{2ex}
\linespread{2.0}
\selectfont
\renewcommand{\thefigure}{S\arabic{figure}}
\setcounter{secnumdepth}{0}
\setcounter{figure}{0}
\tableofcontents
\section{Additional Results}
\subsection{MoTe$_2$ transfer and transient curves}
The transfer curves, I$_{\text{ds}}$-V$_{\text{gs}}$ of two additional devices are shown in Fig. S1(a) and (b). The curves show that different devices, with different resistance and quantitative gate responses show a similar \textbf{qualitative} behavior. For the device shown in Fig. S1(a), the mobility is 0.04 cm$^2$ V$^{-1}$s$^{-1}$ and the sub-threshold swing is $\sim 13$ V dec$^{-1}$, whereas the device in Fig. S1(b) has a mobility of 0.15 cm$^2$ V$^{-1}$s$^{-1}$ and subthreshold swing value similar to the former device. Strikingly, given the large variation in quantitative properties, the trend in behavior, both in the transfer curves and the transient curves (Fig. S2) remains similar in all devices, which is a strong indication to the validity of the proposed model.
\begin{figure}[!htb]
		\label{AddIdVg}
		\centering
		\includegraphics[scale=0.9]{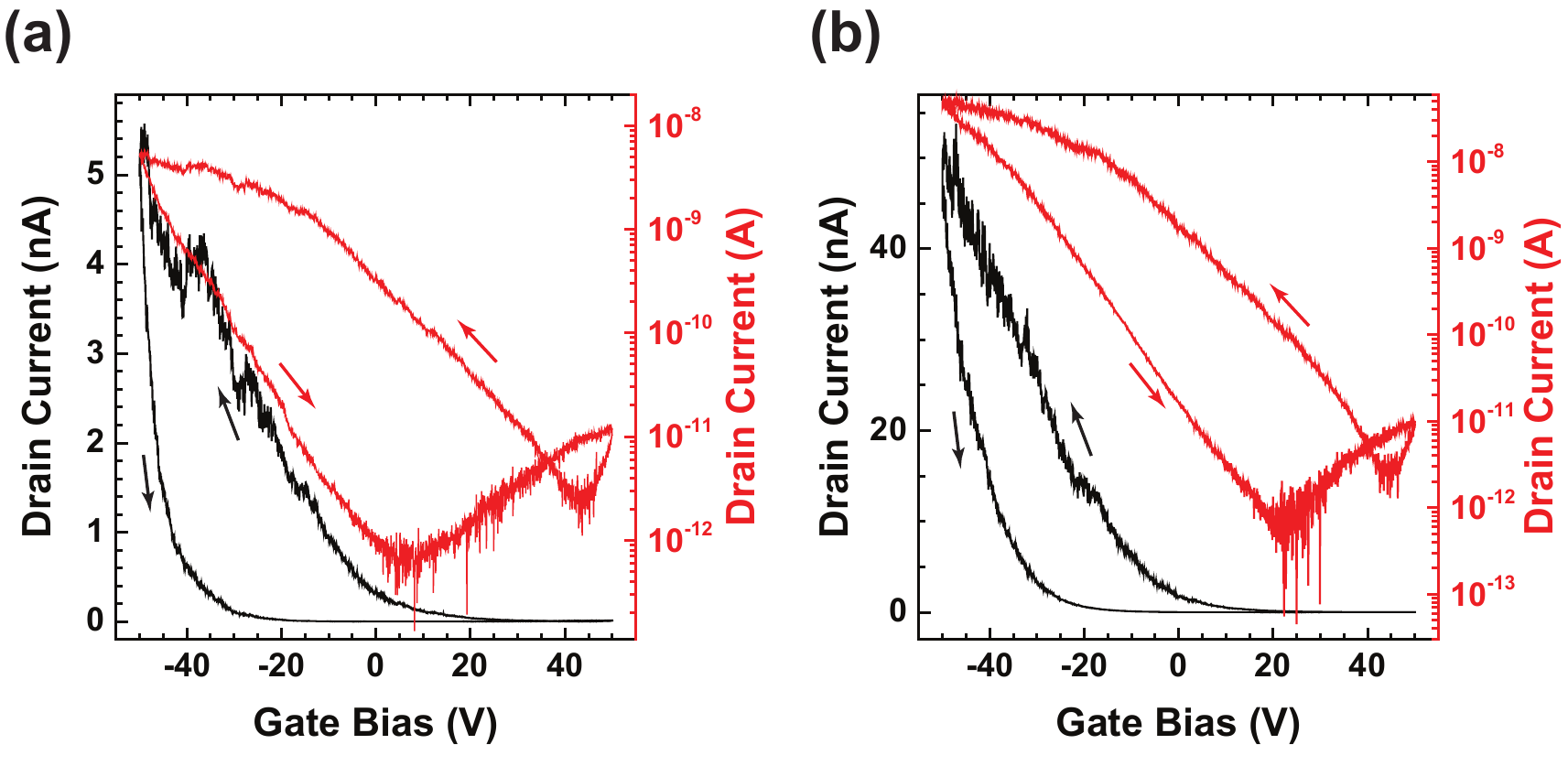}
		\caption{Two additional transfer (I$_{\text{ds}}$-V$_{\text{gs}}$) curves of MoTe$_2$ devices. The black curves are shown on a linear scale and the red curves on a semi-logarithmic scale. Arrows indicate the sweep direction}
	\end{figure}
	
	The transient response to gate pulses was measured on several devices, with different pulse heights. Here we show an analysis done on three additional devices. The results shown in Fig. S1(a) and (d) were collected by pulsing the gate between V$_{\text{gs}} = -10$ V (``open'') and V$_{\text{gs}} = -5$ V (``close''). The results shown for the next two devices were collected using the same dynamic window of $\Delta$V$_{\text{gs}} = 10$ V, as the device discussed in the main text, and the gate was pulsed between V$_{\text{gs}} = -20$ V (``open'') and V$_{\text{gs}} = -10$ V (``close''). The transfer properties of all the MoTe$_2$ transistors are qualitatively comparable.
	
	The transient curves shown in Fig. S1 (a) have a slower rise in the lower values of V$_{\text{ds}}$. Indeed, the pre-exponential decay factor of the deep traps, $A_2$ are constant up to V$_{\text{ds}} = 0.4$ V and sub-linear up until V$_{\text{ds}} = 0.6$ V. This result is attributed to the high conductivity of the channel at V$_{\text{gs}} = -5$ V and the low drift velocity across the channel at V$_{\text{ds}} \leq 0.6$ V. In this combination of parameters the magnitude of the emission current is comparable or larger than the drift current in the channel and, therefore, is the dominant current in the system. However, as the drift velocity increases with increased V$_{\text{ds}}$, the threshold voltage transient becomes more dominant, as is evident from the linear dependence of the pre-exponential decay factor in V$_{\text{ds}}$. 
	
	In contrast, the transient curves shown in Fig. S1 (b) and (c) are of devices with higher resistivity than that of the device discussed in the main text, as is evident from the lower currents. While in these devices, the linear trend (Fig. S1 (e) and (f)) is visible from V$_{\text{ds}} = 0.2$ V onwards, the shape of the transient curves in Fig S1 (b) is quite different from those of the other reported devices. From the analysis, we find that the emission currents from the shallow traps happen at time constants, $\tau _1 \approx 0.4$ s, much smaller than the other reported devices, where $\tau _1$ is in the order of a few tens of seconds. We can, therefore, conclude that in this device, the contribution of the emission currents from the shallow traps were negligible, eliminating the initial (fast) recovery, and thus changing the shape of the transient curves.
	\begin{figure}[!htb]
		\label{AddVdsTrans}
		\centering
		\includegraphics[scale=0.15]{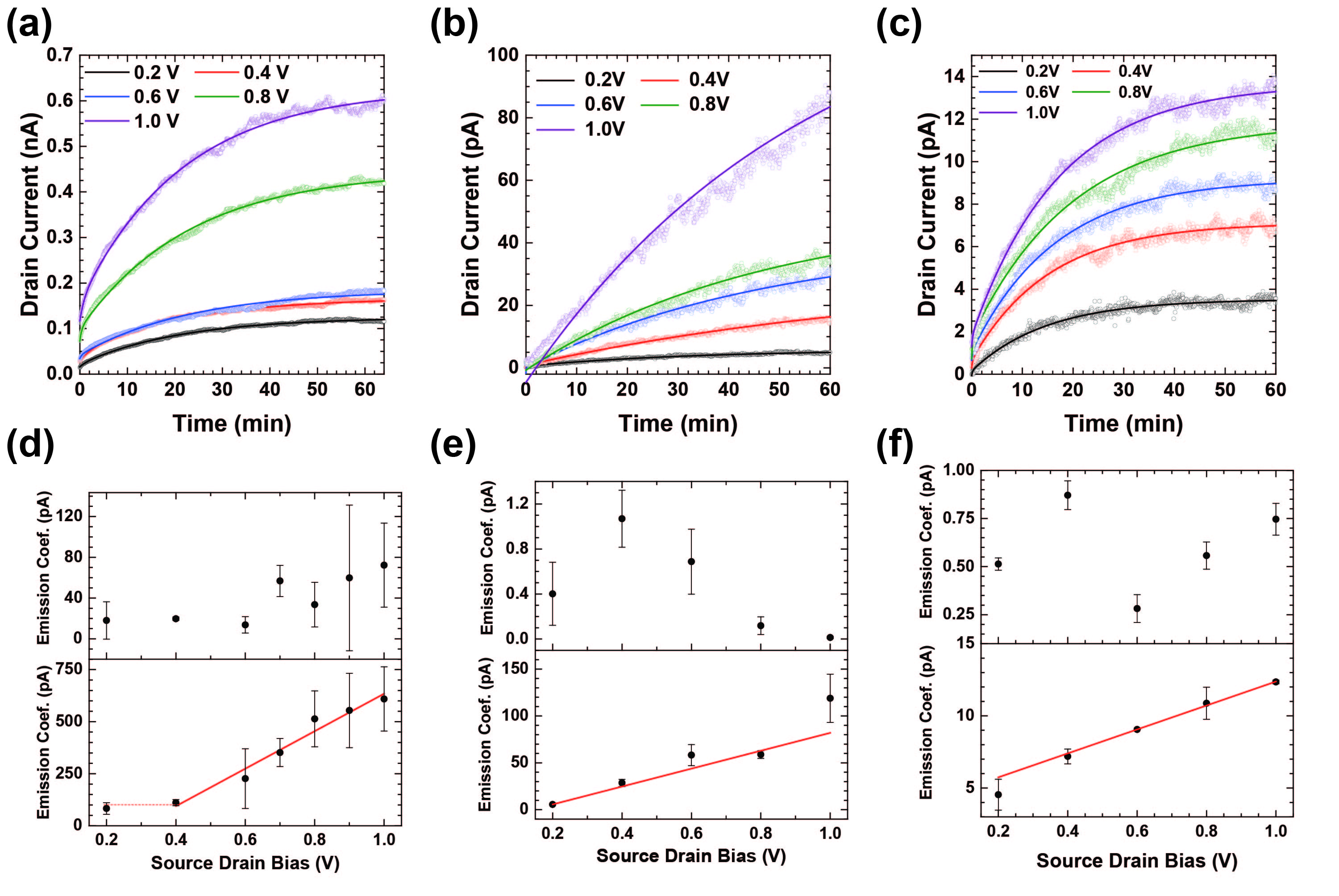}
		\caption{(a)--(c) Current recovery segments, for three different devices, recorded at V$_{\text{gs}} = -5$ V (panel (a)) and V$_{\text{gs}} = -10$ V (panels (b) and (c)) and varying V$_{\text{ds}}$ values, from 0.2V to 1.0V in 0.2V intervals. The circles are the measured data and the solid curves are the double-exponential fits. (d)--(f) The pre-exponential coefficients for the short emission coefficients, $A _1$ (Top Panel) and long emission coefficient $A _2$. Fig S1(d) includes results from V$_{\text{ds}}$ values that are not shown in (a). The red line represents the best linear fit.}
	\end{figure}
	
	\subsection{Threshold voltage transients in WS$_2$}
	The mechanism of threshold voltage transient changes to the conduction of the channel is not limited to devices with high density of trap states. To illustrate the generality of the model, we present some of the threshold voltage transients measured in a CVD grown bilayer WS$_2$ transistor, which was \textit{in-situ} annealed and measured in vacuum ($ < 2\times 10^{-7}$ mbar)
	\begin{figure}
		\label{WS2}
		\centering
		\includegraphics[scale=0.13]{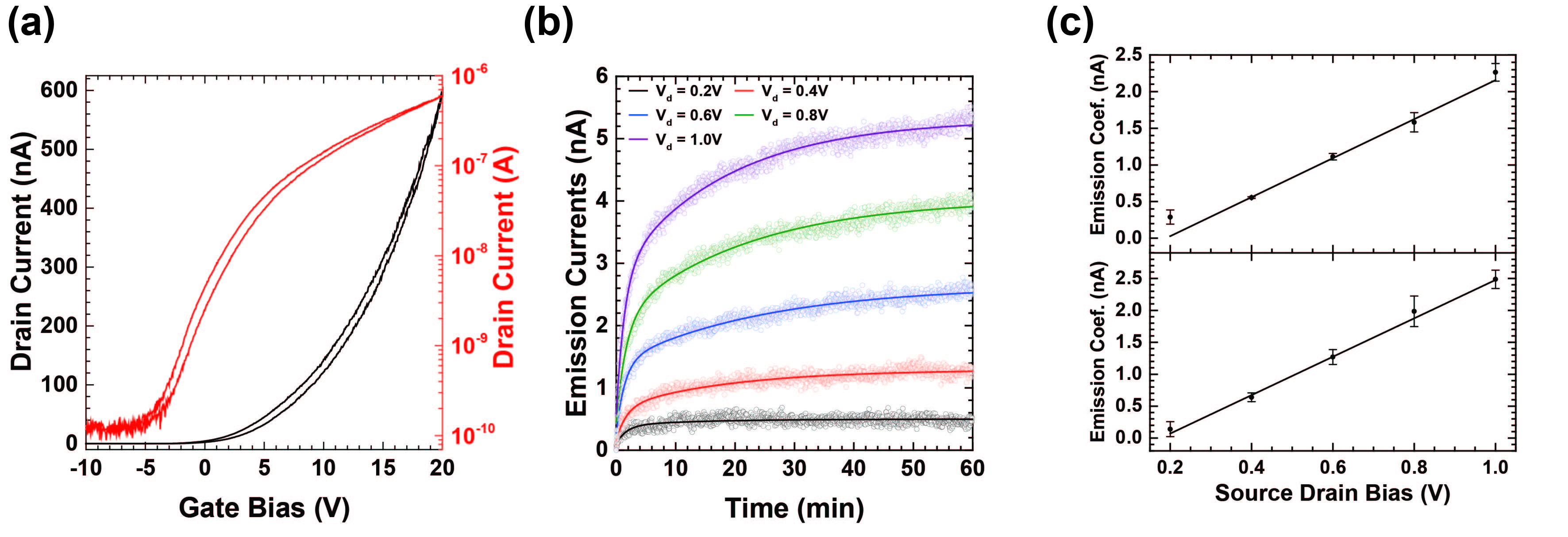}
		\caption{(a) Transfer curves (I$_{\text{ds}}$-V$_{\text{gs}}$) of a CVD-grown WS$_2$ transistor in the linear (black) and semi-logarithmic (red) scales. (b) Current recovery segments recorded at V$_{\text{gs}} = 10$ V and varying V$_{\text{ds}}$ values, from 0.2V to 1.0V in 0.2V intervals. The circles are the measured data and the solid curves are the double-exponential fits. (c) The pre-exponential coefficients for the short emission coefficients, $A _1$ (Top Panel) and long emission coefficient $A _2$. The solid lines represent the best linear fit.}
	\end{figure}
	
	The transfer curve in Fig S3 (a) shows that in the WS$_2$ transistors, the hysteresis is significantly lower than that found in the MoTe$_2$. The transient curves and emission coefficients measurements for WS$_2$ clearly show that the same mechanism of time dependent resistance governs the behavior of the charge conduction in the channel, see Fig. S3 (b) and (c), respectively. Even though the charge carrier mobility in WS$_2$ is more than 100 times larger than that in the studied MoTe$_2$, we still find that both the short lived and the long lived traps contribute to the modulation of channel resistance.
	
	\section{Response Time of the setup}
	To ensure that the measured results are not an artefact of the measurement, the response time of the set-up was measured by pulsing the gate source unit and recording the current through a resistor of comparable resistance to that of the channel ($\sim 1$ G$\Omega$). The rise time of the current was found to be less than 2 $\mu$s, corresponding to the rise time of the current preamplifier. The current follows an exponential rise equation with a rise coefficient $\tau = 0.22$ $\mu$s, thus ensuring that the time response measured o the atomically thin MoTe$_2$ is solely due to the carrier dynamics in the device. Fig. S3 shows the measured rise time of the current.
	\begin{figure}
		\label{TimeResponse}
		\centering
		\includegraphics[scale=0.75]{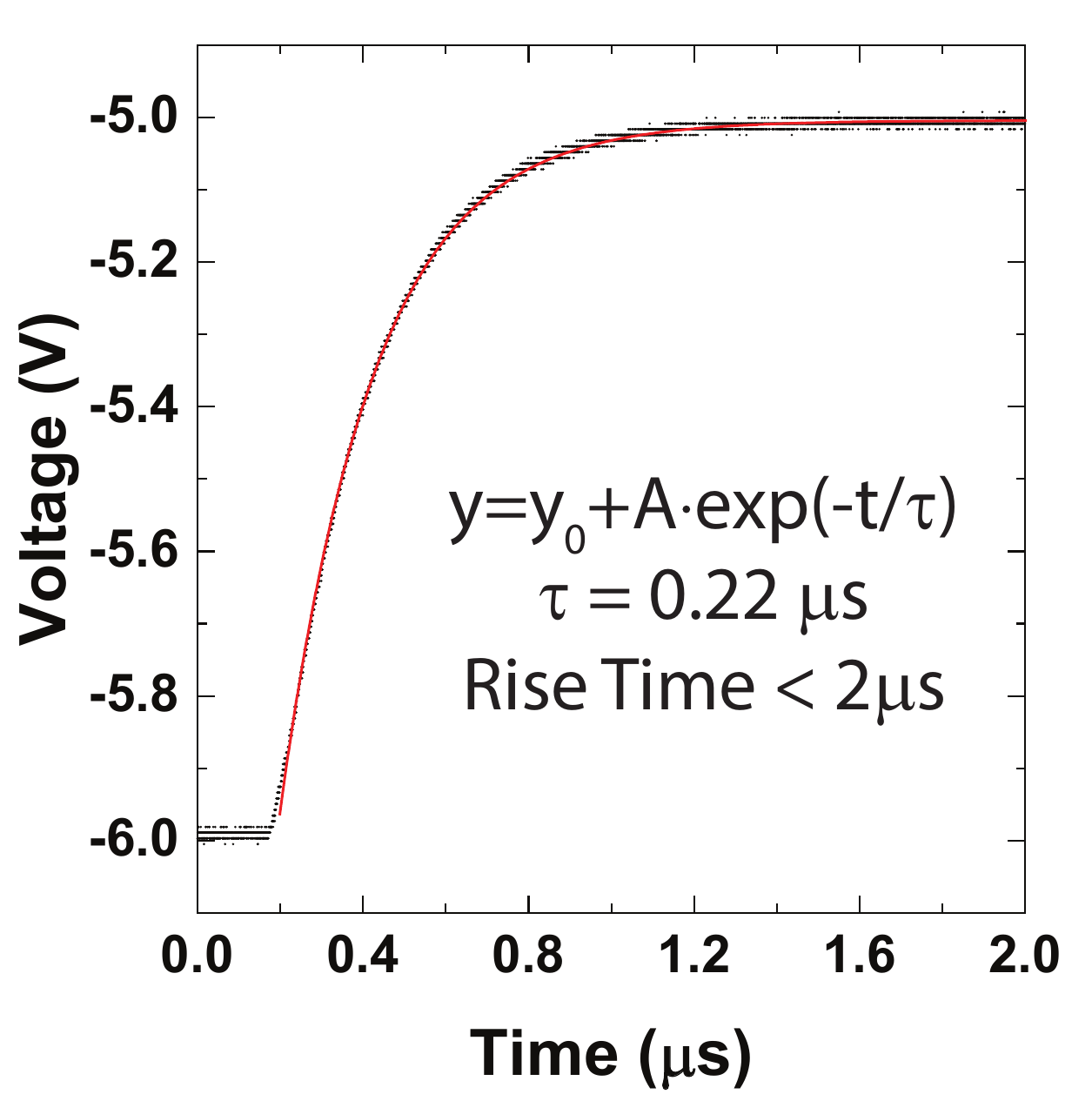}
		\caption{The rise time of the set-up, recorded using a 1 G$\Omega$ resistor}
	\end{figure}
	
	\section{Detailed Derivation of the Threshold Transient Model}
	\paragraph{Classical Current Transient Theory}
	To allow for a thorough discussion of the difference between the classical model of transient currents and the newly presented model which accurately describes the carrier dynamics in atomically thin MoTe$_2$-FETs, we must first present the main aspects of the classical current transients theory. We present an analysis for the process of capturing holes near the valance band maximum ($E_V$), as this is the relevant process for $p$-doped MoTe$_2$. 
	
	The rate of capturing holes from the valance band ($R_{hc}$) is proportional to the density of holes in the valance band ($p$) and the density of unoccupied traps ($N_T - p_T$), where $N_T$ is the total density of trapping states and $p_T$ is the density of occupied states. It's important to note here that ``unoccupied'' from holes means occupied by electrons and electrically neutral. 
	\begin{equation}
	R_{hc} \equiv \left.\frac{\partial p}{\partial t}\right|_{capture} = -c_p (N_T - p_T) p
	\end{equation}
	Where $c_p$ is the capture coefficient for holes, and it equals the thermal velocity, $v_{th}$, multiplied by the capture cross section, $\sigma _p$.
	
	The emission of holes from the traps is described using same considerations without taking into account the unoccupied states in the valance band, since it is assumed that for a non-degenerate semiconductor the emission rate is not limited by it.
	\begin{equation}
	R_{he} \equiv \left. \frac{\partial p}{\partial t}\right| _{emission} = e_p p_T
	\end{equation}
	Where $e_p$ is the emission rate of holes from traps to the valance band. It is therefore clear, that the total change in trap occupation is given by:
	\begin{equation}
	R_p = \left.\frac{\partial p}{\partial t}\right|_{capture} + \left. \frac{\partial p}{\partial t}\right| _{emission} = e_p p_T - c_p (N_T - p_T) p
	\end{equation}
	
	With the traps saturated we can write $N_T = p_T$ and the capture rate will become zero. 
	\begin{equation}\label{emission_rate}
	R_p = \frac{\partial p}{\partial t} = e_p p_T
	\end{equation}
	In a simple process where every hole \emph{added} to the valance band is \emph{removed} from a trap (\textit{i.e.} without any further charge injection), it is clear that:
	\begin{equation}\label{ChargeConservation}
	\frac{\partial p}{\partial t} + \frac{\partial p_T}{\partial t} = 0
	\end{equation}
	Combining Eq. \ref{ChargeConservation} with Eq. \ref{emission_rate} yields
	\begin{equation}\label{Trap_dynamics}
	p_T = p_T(0) e^{-t/\tau}
	\end{equation}
	Where $\tau = 1/e_p$ is the decay constant \emph{per trap}, and $p_T(0) = N_T$ is the trap occupation at the saturation point.
	
	In the classical case, where the entire contribution to the current transient is from charges that are emitted from the traps back into the circuit, the current transient is given by $I(t) = I_0 + qR_p A$, Where $I_0$ is the steady state current, and $A$ is the area from which charges are emitted. Using Eq. \ref{Trap_dynamics}, one can write an explicit expression for the current transient
	\begin{equation}\label{current_transient1}
	\text{I}(t) = \text{I}_0 + \frac{qN_TA}{\tau} e^{-t/\tau}
	\end{equation}
	
	\paragraph{Derivation of the Threshold Voltage Transient} 
	For time-dependent currents that stem from evolution of the threshold voltage in the field-effect transistor (FET), there are a few parameters that determine the current transient. First, the current equation for an FET in the linear regime is
	\begin{equation}\label{current}
	\text{I} _{\text{d}}(t) = \frac{W\mu _p C_{ox}}{L}\left(\text{V} _{\text{th}} (t) - \text{V} _{\text{g}}\right) \text{V} _{\text{d}}
	\end{equation}
	Where $\text{I} _{\text{d}}$ is the drain current, $W$ and $L$ are the channel width and length, respectively, $\mu _p$ is the mobility of the holes, $C_{ox}$ is the capacitance of the gate dielectric, and $\text{V} _{\text{th}}$ and $\text{V} _{\text{g}}$ are the FET threshold voltage for conduction and the gate bias, respectively. The threshold voltage is given by
	\begin{equation}
	\text{V} _{\text{th}} (t) = \Phi _{MS} - \frac{Q_{T} (t)}{C_{ox}} - \Delta E_F
	\end{equation}
	Where $\Phi _{MS}$ is the difference in workfunction between the gate electrode and the conduction channel and $Q_T(t) = \left(Q_0 + qp_T(t)\right)$ accounts for both the stationary charges in the oxide ($Q_0$) and the dynamic charges that are trapped and de-trapped on the channel. It is important to note here that in contrast to a conventional inversion-based FET, the MoTe$_2$ is an accumulation-based transistor. Therefore, the ``textbook'' $2\phi _F$ expression for strong inversion has been substituted here for a general $\Delta E_F$ which represent the change in Fermi energy required to ``open'' the channel. From this equation, one can easily write an expression to describe the dynamics of the threshold voltage
	\begin{equation}
	\frac{d\text{V} _{\text{th}} (t)}{dt} = -\frac{q}{C_{ox}} \frac{dp_T(t)}{dt}
	\end{equation}
	Using a simple model for the concentration of free charge carriers $p(t) = \left(C_{ox} (\text{V} _{\text{th}} (t) - \text{V} _{\text{g}})\right)/q$, it's easy to see that charge is conserved in this model, $p = p_0 - p_T$, where $p_0$ is the total density of holes in the valance band in equilibrium conditions and without traps, and is constant. Using the previously found expression for the emission rate, we can now write an expression for the time-dependent threshold voltage
	\begin{equation}
	\text{V} _{\text{th}} = \Phi _{MS} - \frac{Q_0}{C_{ox}} - \frac{qN_T e^{-t/\tau}}{C_{ox}} - \Delta E_F = \text{V} _{\text{th},sat} - \frac{qN_T e^{-t/\tau}}{C_{ox}}
	\end{equation}
	Where on the right hand side, all the terms that are time-independent were grouped together into $\text{V} _{\text{th},sat}$ The current then becomes
	\begin{equation}
	\text{I} _{\text{d}}(t) = \frac{W\mu _p C_{ox}}{L} \left(\text{V} _{\text{th},sat} - \frac{qN_T e^{-t/\tau}}{C_{ox}} - \text{V} _{\text{g}}\right)\text{V} _{\text{d}} = \text{I} _{\text{d},sat} - \frac{q W\mu _p N_T \text{V} _{\text{d}}}{L} e^{-t/\tau}
	\end{equation}
	
	Finally, to account for the case where the a resistance in series (\textit{e.g.}, contact resistance) plays an important role in the device performance, we add a constant resistance term, $R_S$ to Eq. \ref{current}:
	\begin{equation}
	\text{I} _{\text{d}}(t) = \frac{W\mu _p C_{ox}}{L}\left(\text{V} _{\text{th}} (t) - \text{V} _{\text{g}}\right) \left(\text{V} _{\text{d}} - \text{I} _{\text{d}} R_S\right)
	\end{equation}
	From this equation, we can easily isolate the current term:
	\begin{equation}\label{horror}
	\text{I} _{\text{d}}(t) = \frac{\frac{W\mu _p C_{ox}}{L}\left(\text{V} _{\text{th}} (t) - \text{V} _{\text{g}}\right)}{1 + \frac{W\mu _p C_{ox}}{L}\left(\text{V} _{\text{th}} (t) - \text{V} _{\text{g}}\right) R_S} \text{V} _{\text{d}} 
	\end{equation}
	Which is still linear with V$_{\text{d}}$, in accordance with Ohm's law. The importance of this result is clear when we examine the limits where the contact resistance is much larger than the channel resistance, \textit{i.e.}, when $R_S \gg \left( \frac{W\mu _p C_{ox}}{L}\left(\text{V} _{\text{th}} (t) - \text{V} _{\text{g}}\right)\right)^{-1}$. In this limit, the current simply reduces to $\text{I} _{\text{d}} = \text{V} _{\text{d}} R_S^{-1}$ which is time-independent. On the other limit, $R_S \ll \left( \frac{W\mu _p C_{ox}}{L}\left(\text{V} _{\text{th}} (t) - \text{V} _{\text{g}}\right)\right)^{-1}$, Eq. \ref{horror} simply reduces back to Eq. \ref{current}. The dominant time-dependent characteristics of the emission currents are therefore a strong indication that the major contribution to the transient profile stems from the time-dependent changes in the channel resistance.
\end{document}